\documentstyle[prd,aps]{revtex}
\begin{document}
\draft
\twocolumn[\hsize\textwidth\columnwidth\hsize\csname
@twocolumnfalse\endcsname
 \preprint{SU-ITP-96-52,  hep-th/9611161}
\title{ Black Hole Superpartners and Fixed Scalars}
\author{Renata Kallosh and Andrei Linde}
\address{Physics Department, Stanford University,
Stanford CA 94305-4060, USA}
\date{November 20, 1996}
\maketitle
\begin{abstract}
Some bosonic solutions of supergravities admit Killing spinors of unbroken
supersymmetry. The anti-Killing spinors of broken supersymmetry can be used to
generate the superpartners of stringy black holes. This has a consequent
feedback on the metric and the graviphoton.   We have found however that the
fixed scalars for the black hole superpartners remain the same as for the
original black holes. Possible phenomenological implications of this result are
discussed.
\end{abstract}
\pacs{PACS: 04.65.+e, 04.70.Dy, 11.30.Pb, 11.25.-w \hspace{1.4 cm} SU-ITP-96-52
\hspace {1.4 cm}
hep-th/9611161}
\vskip2pc]

\section{Introduction}
Recently it was found that   black holes serve as attractors for the moduli
fields: The values of the moduli near the horizon do not depend on their values
far away from the black holes \cite{FKS,FK}. For example, in the theory of
dilaton black holes with electric and magnetic charges $q$ and $p$, the dilaton
field   is given by
\begin{equation}\label{1}
 e^{-2\phi (r)} ={  {\mbox{e}}^{-\phi_0} + \frac{|q|}{r} \over
{\mbox{e}}^{+\phi_0}+\frac{|p|}{r}} \ .
\end{equation}
Here $\phi_0$ is the value of the dilaton field at infinity, $r$ is the radial
coordinate which measures the distance from the horizon. The mass $M$ is
related to the electric  and magnetic   charges as
\begin{equation}\label{mass}
M(\phi_0) =  {\textstyle{1\over 2}}({\mbox{e}}^{-\phi_0}  |p| +
{\mbox{e}}^{\phi_0}  |q|)  \, .
\end{equation}
The dilaton field near the horizon  approaches the value
\begin{equation}\label{bhvac1}
e^{-2\phi} = {|q|\over |p|} \
\end{equation}
independently of its value at infinity $\phi_0$. This value corresponds to the
minimum of the mass $M(\phi)$ (\ref{mass}) with respect to $\phi$.
For the non-extreme black holes the situation is similar. The value of
$e^{-2\phi}$ at the horizon is not necessarily i equal to $ {|q|\over |p|} $.
However   the mass of the non-extreme black holes   with fixed electric and
magnetic charges and  fixed entropy $S$  has a minimum at the same point
(\ref{bhvac1}) as the mass of the extreme ones \cite{GKK}.
Finding such a minimum makes operational sense   in the situations where the
black hole evaporation is slow  and  the entropy of a black hole remains
constant.

These results are pretty general; they hold for a wide variety of stringy black
holes \cite{FKS,FK,GKK}. However, it was not known whether the superpartners of
these black holes share the same property. In what follows we will show that
this is indeed the case: superpartners of stringy black holes are classical
solutions with the same electric and magnetic charges, with the same area of
horizon, and with the same values of the moduli fields at the horizon.
Thus the universality of superattractors appears to be even more general than
we expected.

Our results may be helpful for investigation of scattering   on black holes and
their superpartners \cite{SCATT}. On the other hand, they might   provide a
mechanism of moduli fixing in vacuum, and may shed some light on the recently
discovered similarity \cite{K9607} between   spontaneous breaking of N= 2
supersymmetry to N = 1 \cite{n2break_1} and the behavior of scalar fields near
the black hole horizon. The main idea  is based on the observation that black
holes, their antiparticles (black holes with opposite values of magnetic and
electric charges) and their superpartners, which may exist as virtual states in
the vacuum, attract the scalar field $\phi$ to the same point (\ref{bhvac1}).
This may result in the scalar fields being fixed not only near the horizon of
the black holes, but also in the vacuum state containing virtual black holes. A
discussion of this interesting but speculative possibility  will be contained
in Appendix.

\section{Fixed Scalars for Black Hole Superpartners}

The supersymmetric generalizations of the black hole geometries in the context
of pure N=2 supergravity without vector  and hyper multiplets was constructed
starting from the  early 80's \cite{Aich}. By applying long-range N=2
supergauge transformation on Majumdar-Papapetrou configurations, the set of all
superpartners to bosonic multi black holes was exhibited in \cite{Aich,AE86}.

We will study here
what happens with fixed moduli of the bosonic solutions under the
transformations required for the construction of the exact superpartners and
whether moduli are still stable with the account of supergauge transformations.

The short account of the situation in pure N=2 supergravity \cite{Aich,AE86} is
the following\footnote{This  resume follows the one in \cite{BKO} where we have
calculated the norm of the fermionic black hole zero modes.}. The
supersymmetric solutions usually considered have vanishing
fermionic fields. It was explained in \cite{Gibb},  that if non-trivial
supersymmetry parameters with the right regularity
properties exist, one can generate a whole supermultiplet of
solutions.  Solutions   with
non-vanishing fermion fields starting with the purely bosonic ones
has been found for extreme Reissner-Nordstr\"{o}m solutions
  solutions of $N=2$ supergravity in \cite{Aich,AE86} by performing
supersymmetry transformations with the parameters which  converge
asymptotically to global supersymmetry parameters.   The new solutions  are not
gauge-equivalent to the
original ones (it is not possible to go back to bosonic solutions by using
asymptotically vanishing supersymmetry parameters).

There are some non-trivial
supersymmetry transformations generated by the Killing
spinors of unbroken supersymmetry
that leave the original bosonic solution invariant.  Only the
supersymmetry parameters corresponding to the broken supersymmetries,
which  we will called   ``anti-Killing spinors" in \cite{BKO}, \cite{BHGAUGE},
generate new non-gauge equivalent solutions.

Consider a  solution of the Einstein--Maxwell theory
 for zero magnetic and positive electric
charge
\begin{equation}
ds^2~=~ V^{-2} dt^2 ~-~ V^2 d{\vec x}^2,\quad
A_t dt ~=~ -{1\over{\kappa}} V^{-1}~dt\ \ ,
\end{equation}

\noindent where $\kappa^{2}=4\pi G$ and the function $V$ is
\begin{equation}
V(\vec{x})\ =\ 1\ +\frac{G M}{|\vec x\ |}\  ,
\label{V}
\end{equation}

\noindent where the horizon of the $s$th black hole is located at
$|\vec{x}|=0$ and its electric charge is
$Q_{s}=G\kappa^{-1}M_{s}$.
This background  admits $N=2$
supergravity Killing spinors \cite{Gibb}, {\it i.e.}~a
solution
of the equation

\begin{equation}
\hat\nabla_\mu
\epsilon_{(k)}~= (\nabla_\mu -\frac{1}{4} \kappa\ {{\slash
\hskip -7.5pt}{} F}\gamma_\mu\ ) \epsilon_{(k)} = 0  \ ,
\end{equation}
where ${{\slash \hskip -7.5pt}{}
F}=\gamma^{\mu\nu} F_{\mu\nu}$ and $F_{\mu\nu}$ is the
field--strength
of the gauge field $A_\mu$ and $\epsilon$ is a Dirac spinor.  That
solution
is given by
\begin{equation}
\epsilon_{(k)} ~=~ V^{-1/2}{\cal C}_{(k)}  \ ,
\end{equation}
 where ${\cal C}_{(k)}$ is a constant spinor satisfying the
condition
\begin{equation}
\gamma_0\ {\cal C}_{(k)} = +\ {\cal C}_{(k)}  , \qquad  {\cal C}_{(k)}{} =
\left (\matrix{
c\cr
c\cr
}\right ),
\label{property}
\end{equation}
 and is given in terms of a complex two-component spinor,
$c$. This means that the
background
given above has one unbroken supersymmetry in $N=2$ supergravity.
The asymptotically constant anti-Killing spinor can be chosen to be, in terms
of
the Killing spinor,
\begin{equation}
\epsilon^{(\bar k)} = -i\gamma_5 \epsilon_{(k)} =  V^{-1/2}{\cal C}^{(\bar k)}\
 ,
\label{anti-Killing}
\end{equation}
and
\begin{equation}
\gamma_0\ {\cal C}^{(\bar k)} = -\ {\cal C}^{(\bar k)} \ ,  \qquad {\cal
C}^{(\bar k)} ~=~  {c\choose{-c}}.
\label{antiproperty}
\end{equation}
It represents broken supersymmetry of the bosonic solution and is used to
generate gravitino
 \cite{Aich}:
\begin{equation}
\psi_\mu~=~ {1\over{\kappa}} \hat\nabla_\mu\epsilon^{(\bar k)}\ \ .
\label{GRAV0M}
\end{equation}
The explicit expression for the gravitino, which solves the field eqs. for
gravitino and  which is linear in anti-Killing spinor is
\begin{equation}
\label{n=2grav}
\psi ~=~ {1\over{\kappa}} V^{-7/2} \partial_{\underline{i}} V
\gamma_i
{\cal C}^{(\bar k)}~dt~+~
{1\over{\kappa}} V^{-3/2} \partial_{\underline{j}} V \gamma_j
\gamma_i
{\cal C}^{(\bar k)}~dx^{\underline{i}}\ \ . \nonumber \\ \relax
\end{equation}
Having found gravitino in the linear order (\ref{n=2grav}) one proceeds with
the iterative procedure and calculates the feedback of the gravitino on the
geometry and the vector field via the supersymmetry transformation generated by
the anti-Killing spinor:
\begin{eqnarray}
\delta e_{\mu}{}^a &=& -{i \kappa \over 2} \left (\bar \epsilon^{(\bar k)}
\gamma^a \psi_\mu - \bar  \psi_\mu \gamma^a \epsilon^{(\bar k)}\right)
 , \nonumber \\
\delta A_\mu &=& {i\over 2} \left (\bar \epsilon^{(\bar k)}   \psi_\mu - \bar
\psi_\mu  \epsilon^{(\bar k)}\right) .
\end{eqnarray}
This leads to the corrections to the original bosonic solution which is of the
second order in Grassmann parameter $c$. This in turn induces additional
corrections to the gravitino etc.
The full procedure leads to an exact solutions of the full supergravity
equations of motion and the series stops after the fourth order in Grassmann
numbers. The actual computation was performed in \cite{AE86} with the help of
the algebraic computer program REDUCE. The result schematically can be
represented in the form
\begin{eqnarray}
g_{\mu\nu}(\vec x, c ) &=&g_{\mu\nu} (\vec x  )+ (c^\dagger  \sigma_i c)\,
\Delta^i_{\mu\nu} (g) + (c^\dagger c)  (c^\dagger c)\,  \Delta_{\mu\nu} (g) ,
\\
\nonumber\\
A_{\mu} (\vec x, c )&=&A_{\mu} (\vec x ) + (c^\dagger  \sigma_i c)\,
\Delta^i_{\mu} (A) +(c^\dagger c)  (c^\dagger c)\,  \Delta_{\mu} (A),
\end{eqnarray}
where the explicit form of the terms quadratic and quartic in Grassmann numbers
$c$ can be found in eqs. (2.34)-(2.36) of the first ref. in \cite{AE86}.
Even when one considers only the near horizon geometry, one still finds that
the superpartners have corrections, e.g, the non-diagonal term in the metric as
well as the space component of the vector which are proportional to
$(c^\dagger  \sigma_i c)$ are present, as different from the original geometry
which was diagonal and from the vector field which had no space-time component.
 Also the non-diagonal terms in the 3-dimensional geometry in the quartic order
in Grassmann variables are present as different from the original geometry.

The generalization of the procedure \cite{Aich,AE86} of the generating exact
superpartners of black holes of  N=2 supergravity interacting  many vector
multiplets and hypermultiplets, in principle, can be performed and may also
require a considerable effort and most certainly will require the
help of  a computer. However for our purpose to understand whether the fixed
scalars will be affected and in which way there is a  shortcut and we may solve
completely the problem with respect to double-extreme black holes \cite{KSW}.
This will be sufficient to understand
the effect of the superpartners of any black holes near the horizon, since all
supersymmetric  black holes near the horizon  tend to the double-extreme ones.
The crucial point here comes when one looks into the gaugino and hyperino
supersymmetric transformation rules. In the first approximation the fermionic
fields are absent and we
have \footnote{In notation of \cite{ABCDFFM}, \cite{FK}. We skip the spinorial
index  $A$ in  for simplicity.}:
\begin{eqnarray}
\delta \psi_{\mu} &=& {\cal D}_\mu \epsilon + T_{\mu\nu}^- \gamma^\nu \epsilon
\equiv  \hat {\cal D}_\mu \epsilon \ , \nonumber\\
\delta \lambda^{i} &=& i\gamma^\mu \partial_\mu z^i \epsilon+
{i\over2} {\cal F}^{i-}_{\mu\nu}\gamma^{\mu\nu}\epsilon \ , \nonumber\\
\delta \zeta_\alpha &=& i \,{\cal U}^{ \beta} _u \partial_\mu  q^u \gamma^\mu
\epsilon
 C_{\alpha \beta} \ ,
\label{ricca}
\end{eqnarray}
where $\lambda^{i}$, $\psi_{\mu}$ are the chiral gaugino and gravitino
fields, $\zeta_\alpha$ is a hyperino. We consider here as  in \cite{FK} the
scalars $q^u$ from the hypermultiplets  to be constant for  the simplest
supersymmetric black hole.  The value of this constant does not affect the
black hole solution since the scalars from the hypermultiplets do not couple to
the vectors. Thus for the constant quaternionic scalars $q^u$ the condition of
unbroken supersymmetry
\begin{equation}
\delta \zeta_\alpha = i \,{\cal U}^{ \beta} _u \partial_\mu  q^u \gamma^\mu
\epsilon
 C_{\alpha \beta}=0 \
\end{equation}
 is satisfied without any constraint on the supersymmetry parameters
$\epsilon$.
The vector multiplets include gaugino, Kahler moduli $z^i$ and the vector
fields ${\cal F}^{i-}_{\mu\nu}$.
Double-extreme black holes have everywhere constant moduli $\partial_\mu
z^i=0$and vanishing  vector field strength ${\cal F}^{i-}_{\mu\nu}=0$. This
last equation actually means that the central charge is extremized in the
moduli space and that the moduli become fixed functions of charges. Thus  the
unbroken supersymmetry equations for gaugino,
\begin{equation}
\delta \lambda^{i} = i\gamma^\mu \partial_\mu z^i \epsilon+
{i\over2} {\cal F}^{i-}_{\mu\nu}\gamma^{\mu\nu}\epsilon =0 \ ,
\end{equation}
are satisfied   without any constraint on the supersymmetry parameters
$\epsilon$.
The gravitino supersymmetry transformations even for the double-extreme black
hole vanishes only with the Killing spinor satisfying the linear constraint
\begin{equation}
\delta \psi_{\mu} = {\cal D}_\mu \epsilon_{ (k)} + T_{\mu\nu}^- \gamma^\nu
\epsilon_{(k)} =0 \ ,
\end{equation}
since the geometry is that of the extreme Reissner-Nordstrom type with the
area-mass formula defined by the charges of the theory. Now we have got enough
information to generate the double-extreme black hole superpartners.
The first corrections to fermions is given by the anti-Killing spinor:
\begin{eqnarray}\label{sp}
\delta^{(1)} \psi_{\mu} &=& \hat {\cal D}_\mu \epsilon^{(\bar k)}  \ ,
\nonumber\\
\delta^{(1)} \lambda^{i} &=& i\gamma^\mu \partial_\mu z^i \epsilon^{(\bar k)} +
{i\over2} {\cal F}^{i-}_{\mu\nu}\gamma^{\mu\nu}\epsilon^{(\bar k)}=0  \ , \\
\delta^{(1)} \zeta_\alpha &=& i \,{\cal U}^{ \beta} _u \partial_\mu  q^u
\gamma^\mu \epsilon^{(\bar k)}
 C_{\alpha \beta}=0  \ . \nonumber
\end{eqnarray}
The absence of corrections to gaugino and hyperino for the double-extreme black
holes follows from their property
\begin{equation}
\partial_\mu z^i=0\ , \qquad {\cal F}^{i-}_{\mu\nu}=0\ , \qquad  \partial_\mu
q^u=0 \ .
\end{equation}

Thus if one starts with the double-extreme bosonic black hole and performs a
long-range N=2 supergauge transformation on fermions, only the gravitino field
is generated, the gaugino as well as the hyperino do not appear even in
presence of Grassmann numbers $c$,
in terms of which gravitino is linear. The complete form of supersymmetry
transformation on bosons  is
\begin{eqnarray}
\delta e_\mu{}^a &=& -i  \bar \epsilon^{(\bar k)}  \gamma^a \psi_\mu + c.c. \ ,
\\
\delta A_\mu ^{\Lambda}&=& 2\bar L^\Lambda \,\bar\psi_{\mu}\, \epsilon   +  i
f^\Lambda_i\, \bar \lambda^{i}\, \gamma_\mu\, \epsilon +  c.c. \ ,\\
\delta z^i &=& \bar \lambda^{i}\, \epsilon \ ,\\
\delta q^u &=& {\cal U}^{u} _{\alpha }\Bigl(\bar\zeta^\alpha \epsilon +
C_{\alpha \beta}  \bar\zeta_\beta\epsilon \Bigr) \ .
\label{bosons}\end{eqnarray}
In the first approximation using (\ref{sp}) one gets the corrections to bosons
which are second order in Grassmann numbers:
\begin{eqnarray}
\delta^{(2)} e_\mu{}^a &=& -i  \bar \epsilon^{(\bar k)}  \gamma^a \delta^{(1)}
\psi_\mu + c.c. \ , \\
\delta^{(2)} A_\mu ^{\Lambda}&=& 2\bar L^\Lambda \,  \delta^{(1)}
\bar\psi_{\mu}\, \epsilon^{(\bar k)} + c.c.  \ , \\
\delta^{(2)} z^i &=& 0 \ ,\\
\delta^{(2)} q^u &=& 0\ .
\end{eqnarray}
The virbeins and the metric transform via gravitino and will get corrections of
the type found in pure supergravity. The vector fields $A_\mu ^{\Lambda}$  get
corrections via gravitino. It is important to check what kind of corrections
the graviphoton
$$T_{\mu\nu}^- = 2i Im {\cal N}_{\Lambda \Sigma} L^\Lambda F^{-\Sigma}$$
has and what happens with the vector fields which are partners of gaugino
$${\cal F}^{i-}_{\mu\nu}= 2iG^{i\bar j}  Im {\cal N}_{\Lambda \Sigma} \bar
f^\Lambda_{\bar j} F^{-\Sigma}  \ . $$
Here one has to remember that the graviphoton and the vectors ${\cal
F}^{i-}_{\mu\nu}$ are orthogonal combinations. The important identity of
special geometry explaining this can be found in eq. (54) of \cite{CDF}.
\begin{equation}
Im {\cal N}_{\Lambda \Sigma} \bar f^\Lambda_{\bar j} L^\Sigma=0 \ .
\end{equation}
Using this we find that only the graviphoton has first order corrections
quadratic in Grassmann numbers and the partner of gaugino has no such
corrections. This is consistent with the fact that our fixed scalars $z^i$
remain fixed in this order. Thus we get second order corrections for the bosons
in the supergravity multiplet
\begin{eqnarray}
\delta^{(2)} e_\mu{}^a &=& -i  \bar \epsilon^{(\bar k)}  \gamma^a \hat
\nabla_\mu \epsilon^{(\bar k)} + c.c. \ , \nonumber \\
\delta^{(2)}  T_{\mu\nu}^- &=& 2 i Im {\cal N}_{\Lambda \Sigma} L^\Lambda
\delta^{(2)} F^{-\Sigma}  \ ,
\end{eqnarray}
and no second order corrections for the bosons  in the vector and hyper
multiplets.
\begin{eqnarray}
\delta^{(2)}  {\cal F}^{i-}_{\mu\nu} &=& 0 \ , \nonumber\\
\delta^{(2)} z^i &=& 0 \ ,\\
\delta^{(2)}  q^u &=& 0 \ . \nonumber
\end{eqnarray}
To solve the problem in the next order we have to look back into the gaugino
and hyperino transformation and take into account the presence of fermions and
fermionic corrections to bosons. This will give correction  terms cubic in
Grassmann numbers for fermions. The complete  form of the fermionic  susy
transformations is rather complicated and can be read off from eqs. (8.24)-
(8.41) of  Ref. \cite{ABCDFFM}.
By carefully checking all terms in the full supersymmetry transformation rules
we find that the gravitino indeed has corrections of the third order in
Grassmann numbers, but neither gaugino nor hyperino have any corrections
\begin{eqnarray}\label{3}
\delta^{(3)} \psi_{\mu} &\neq & 0\ , \nonumber\\
\delta^{(3)}  \lambda^{i} &=& 0\ ,  \\
\delta^{(3)} \zeta_\alpha &=& 0\ .  \nonumber
\end{eqnarray}
This we can plug back one more time into the bosonic supersymmetry
transformations (\ref{bosons}) and we get the result we looked for: there are
4-th order terms in the bosons of the supergravity multiplet of the structure
displayed in \cite{Aich,AE86}
\begin{eqnarray}
\delta^{(4)} e_\mu{}^a &\neq & 0\ , \nonumber\\
\delta^{(4)}  T_{\mu\nu}^- &\neq & 0  \ ,
\end{eqnarray}
and there are no 4-th order corrections for the bosons  in the vector and hyper
multiplets:
\begin{eqnarray}
\delta^{(4)}  {\cal F}^{i-}_{\mu\nu} &=& 0 \ , \nonumber  \\
\delta^{(4)} z^i &=& 0 \ ,\\
\delta^{(4)}  q^u &=& 0 \ \nonumber  .
\end{eqnarray}
Thus indeed the fixed point structure of the near horizon configuration is
stable to the supersymmetry transformations which generate the fermionic
partners of the supersymmetric black holes.  Although the metric and the
graviphoton have corrections of the second and fourth order in Grassmann
variables and gravitino has terms of the third order, nothing happens with the
fixed scalars and their fermionic partners: there are no corrections in all
possible orders in Grassmann variables. In particular this means that the
moduli $z^i$ in the vector multiplets remain the same functions of charges near
the horizon  $z^i_{\rm fix} (p,q) $ for the black hole superpartners as they
were for the original black holes.

\section{Discussion}

In this paper we have shown that the same mechanism which fixes the scalar
fields near the black hole horizon works for the superpartners of the black
holes as well. Black holes and their superpartners have the same mass, charges,
area of horizon, and entropy. In addition, the scalar fields near the horizon
of black holes and of their superpartners reach the same value at the horizon.
Thus we have found an interesting universality of  the  properties of all
members of the black hole hypermultiplet.

This result may have many interesting implications in quantum theory of black
holes. We will argue in Appendix   that virtual black holes and their
superpartners may fix the values of the moduli fields in the vacuum. Whereas
this idea is very speculative (that is why it is in Appendix), it deserves
further investigation because it may provide a link between the supersymmetric
models with symmetry breaking N = 2 $\to $ N = 1 and black hole physics.

\section*{Acknowledgements}
The authors appreciate very stimulating discussions with T. Banks, M. Dine, G.
Gibbons, S.-Y. Rey, A. Strominger, and L. Susskind.
This
work was  supported
by  NSF grant PHY-9219345.

\section*{Appendix. Is it possible to fix scalar fields by virtual black
holes?}

As we have found, not only black holes but their superpartners as well fix  the
values of the scalar fields on the horizon. Moreover, the masses of the black
holes with given values of magnetic and electric charges become minimal if the
value of  the scalar field were equal to its value at the horizon in the whole
universe. This particular value minimizes the black hole mass even if it is not
extremal \cite{GKK}.

This makes it very tempting to consider only those configurations for which the
total energy (mass) is minimal, and the scalar fields take the same values at
the horizon and at infinity. Such configurations are called double-extreme
\cite{KSW}. They are particularly simple and allow  much more detailed
investigation than the black holes with scalar fields changing with the
distance from a black hole.

Let us consider again the simplest example described in the Introduction. The
scalar field (\ref{1}) near the horizon approaches the asymptotic value
(\ref{bhvac1}),
$e^{-2\phi} = {|q|\over |p|}$.
This asymptotic regime is reached only very close to the horizon. For example,
for $p,q,\phi_0 \sim 1$ the dilaton stabilization occurs only at the Planckian
distance from the horizon, $r < 1$; for larger $p$ and $q$ the asymptotic
regime occurs earlier, at greater $r$.

The same property is shared by the multi-black-hole solutions describing a
collection of  extreme black holes with charges $cq$ and $cp$, where $c$ is any
constant: Near each of the black holes the scalar field will approach the same
value given by (\ref{bhvac1}). If the black holes are very close to each other
in some part of space, then the value of the dilaton field in this region will
be close to (\ref{bhvac1}) independently of $\phi_0$.

But do we actually need the gas of real black holes, or would it be enough to
consider virtual black holes appearing in the vacuum and disappearing again? An
important feature of the black hole attractors which is manifest in the
simplest case considered in the Introduction is that the value of the dilaton
field $\phi$ near the  horizon of an extreme black hole  does not depend on the
sign of $q$ and $p$. It is also important  that the masses of black holes
(either extreme or non-extreme) have a minimum at the same value of $\phi$,
independently of the sign of  $q$ and $p$. Therefore both black holes and
anti-black holes (black holes with opposite charges) will push the dilaton
field to the same point (\ref{bhvac1}), corresponding to the minimum of the
effective mass $M(\phi)$ (\ref{mass}).  Finally, in the previous section we
have found that superpartners attract the dilaton field towards the same point.
This makes it very tempting to speculate that   black holes, their
anti-particles and superpartners which exist as virtual states in the vacuum,
may stabilize the values of moduli field in the vacuum, even in the absence of
real black holes.

To make our idea more clear,   suppose first that the only black holes that may
appear in the vacuum are the ones with charges $\pm  p$ and $\pm  q$.   Let us
assume for a moment that their contribution to the effective potential is given
by the standard one-loop expression for bose particles:
\begin{equation}\label{bhV}
V(\phi) \sim \int d^4k\, \ln[k^2 + M^2(\phi)] \sim \int \limits_{0}^{\Lambda}
dk\, k^3 \ln[k^2 + M^2(\phi)] \ .
\end{equation}
Here $\Lambda$ is the ultraviolet cut-off. Normally in quantum field theory one
takes the limit $\Lambda \gg M$, calculates the integrals and makes the
renormalization if necessary. In our case the situation is not that simple. In
string theory  there may be no momenta greater than $M_p = 1$. Meanwhile the
mass of a black hole with large $p$ and $q$ is much greater than $1$. Therefore
the integral in (\ref{bhV}) for large black holes should be calculated in a
rather unusual limit $\Lambda \sim 1 \ll M$. This gives the following estimate
for the dilaton effective potential induced by virtual black holes \cite{Rey}:
\begin{equation}\label{bhV2}
V(\phi) \sim     \int \limits_{0}^{1} dk\, k^3 \ln[k^2 + M^2(\phi)]\ \sim \
\ln M^2(\phi)   \ .
\end{equation}
Obviously, this effective potential has a minimum at the same point as the
black hole mass $M(\phi)$ (\ref{mass}). Thus if our arguments are correct,
nonperturbative effects associated with virtual black holes may stabilize the
dilaton field at the same place at which the dilaton field is stabilized near
the black hole horizon.

There are several problems associated with this proposal. Eq. (\ref{bhV}) was
written by analogy with the theory of point-like bose particles.  However,
black holes   with non-vanishing entropy  $S = 2\pi |pq|$ are not point-like
objects, they are solitons which have finite size, and finite area of horizon
and finite entropy. For large $p$ and $q$ they are surrounded by a  large
classical dilaton field, which in a certain sense can be considered a
collection of many bose particles in the same state. If  black holes were
normal point-like particles described by quantum field theory,   their
contribution to $V(\phi)$  would be exactly cancelled by the contributions of
their superpartners. One may expect that supersymmetry transformation
transfers one of these bosons into a fermion. However, this does not
necessarily mean that the contribution of such states to the effective
potential must cancel the contribution of the original black holes carrying  no
fermions. To answer this question one should  perform a real calculation of
nonperturbative quantum effects induced by virtual black holes. In particular,
one should take into account that non-extreme black holes may give a
contribution to $V(\phi)$ as well, and this contribution will not be cancelled
by the contribution of their superpartners because non-extreme black holes
break supersymmetry. As we already emphasized, the mass $M(\phi)$ of
non-extreme black holes with given $p, q$ and the entropy $S$ has a minimum
with respect to $\phi$ at the same point as the  mass of extreme black holes
with the same $p$ and $q$. Therefore one may expect that if one takes the
contribution of non-extreme black holes with the same absolute values of $p$
and $q$ and with the same entropy $S$, the resulting contribution will not be
cancelled by the contribution of their superpartners. In this respect it is
encouraging that the arguments presented above can get an additional support
from our results concerning black hole superpartners. If one tries to visualize
the vacuum state as a medium populated by virtual black holes and their
superpartners, then from our results  it follows that not only black holes but
their superpartners as well  attract the scalar fields to the same values. This
makes it more plausible that virtual black holes and their superpartners may be
responsible for fixing the values of the moduli fields in the vacuum.

On the other hand, the same reason which may produce a nonvanishing
contribution of black holes to the effective potential (the non-perturbative
nature of this effect) may imply that the effect in fact will be exponentially
suppressed, just like any effects in the theory of instantons. Using
thermodynamical analogy described in \cite{US}, one may expect that the
``abundance'' of virtual black holes in the vacuum should be suppressed by an
exponential factor of the type of $e^{-2S} = e^{-4\pi |pq|}$ describing
suppression of probability of a simultaneous production of a pair of black
holes. This factor  should be included in  (\ref{bhV}):
\begin{equation}\label{bhV3}
V(\phi) \sim   {e^{-2 S(p,q)} }\, F(M^2(\phi)) \ .
\end{equation}
Here we have written some function $F(M^2(\phi))$ instead of $\ln M^2(\phi)$ to
reflect the uncertainty of the estimates described above. We should emphasize
that we used Coleman-Weinberg approach  only as a heuristic way to estimate
the effective potential $V(\phi)$.  Fortunately, for the validity of our
argument we do not really need  to know the function $F(M^2(\phi))$ exactly.
The only thing which is  important to us is that if $F(M^2(\phi))$  is a
monotonous function  such as  $M^2(\phi)$ or $\ln M^2(\phi)$ (\ref{bhV2}), then
it has a minimum at the same point where $M^2(\phi)$ has a minimum, i.e. at
$e^{-2\phi} = {|q|\over |p|}$, see eq. (\ref{bhvac1}). Thus one may argue that
virtual extreme and non-extreme black holes can fix  the dilaton field in
vacuum at the same value at which   extreme black holes fix this field near the
horizon.

In the models describing N = 4 axion-dilaton black holes with electric and
magnetic charges $n_1, n_2, m_1, m_2$  \cite{K9607} the values  of the dilaton
field $\phi$ and the axion field $a$ at the extreme black hole horizon are
given by
\begin{equation}
e^{-2\phi} = {|n_2\, m_1 - n_1\, m_2| \over m_1^2 + m_2^2}\ , \qquad
 a  = {n_2 m_2 + n_1 m_1 \over m_1^2 + m_2^2} \  .
\label{attr}\end{equation}
 As a result, one may expect that the sum of contributions of all virtual black
holes  with absolute values of electric and magnetic charges $|n_1|, |n_2|,
|m_1|, |m_2|$  fixes the value  of the   dilaton field   at the   point
(\ref{attr}). Note that this point  which for the case $m_2= 0$ corresponds to
the minimum of the dilaton potential in the supersymmetric model of ref.
\cite{n2break_1}. This suggests that virtual black holes may provide a
dynamical mechanism explaining a mysterious correspondence \cite{K9607} between
the behavior of scalar fields in new models of breaking of N = 2 SUSY down  to
N = 1 \cite {n2break_1} and the properties of scalar  fields near the horizon
of the axion-dilaton black holes.

If one simultaneously changes sign of $n_i$ and $m_i$ (which should be allowed
because we consider contributions of black holes and their anti-particles) the
value of the axion field in (\ref{attr}) does not change. However, if there
exist black holes with charges $(n_i, - m_i)$, then the field $a$ near the
horizon of such black holes will change its sign. Therefore the possibility to
fix the axion field in the vacuum requires a more careful analysis.

In a more general case one may consider vacuum   populated   by black holes
belonging to several different   hypermultiplets which are compatible with a
given mechanism of stringy compactification. All such hypermultiplets may give
contributions to the effective potential $V(\phi)$. Thus all of them should be
taken into account, and one may need to take a sum over contributions of black
holes with all possible values of $p$ and $q$. To find a complete expression
for the effective potential one would need to perform  a detailed calculation
involving virtual black holes and their superpartners. Perhaps it would be more
appropriate   to study this issue in the context of D-brane theory. However,
the calculation of the effective potential in this theory would go far beyond
the scope of this paper.  As a first step in this direction, one should
establish relation between D-branes and fixed scalars. This will be a topic of
a separate publication \cite{KallD}.


\begin{references}

\bibitem{FKS}
S. Ferrara,  R. Kallosh, and A. Strominger,    Phys. Rev.  D {\bf 52},
5412,1995; hep-th/9508072.
\bibitem{FK} S. Ferrara and R. Kallosh,   Phys.Rev. D {\bf 54}, 1514 (1996),
hep-th/9602136; Phys.Rev. D {\bf 54}, 1525 (1996),
 hep-th/9603090.
\bibitem{GKK} G. Gibbons, R. Kallosh, and B. Kol,  hep-th/9607108.
\bibitem{SCATT} B. Kol and A. Rajaraman,   hep-th/9608126; S.R. Das, G.W.
Gibbons, and S.D. Mathur,   hep-th/9609052; C.G. Callan,   S.S. Gubser, I.R.
Klebanov, and  A.A. Tseytlin, hep-th/9610172.
\bibitem{K9607} R. Kallosh, Phys.Rev.  D {\bf 54},  4709 (1996),
hep-th/9606093.
\bibitem{n2break_1} S.~Ferrara, L.~Girardello, and M.~Porrati, Phys.  Lett.
{\bf B366}, 155 (1996), hep-th/9510074; I. Antoniadis, H. Partouche, and T.R.
Taylor, Phys. Lett. {\bf B372}, 83 (1996);
 I. Antoniadis and T.R. Taylor,   hep-th/9604062; S. Ferrara, L. Girardello,
and M. Porrati, Phys. Lett. {\bf B376}, 275 (1996). hep-th/9512180.
\bibitem{Aich}{P.C.~Aichelburg and R.~G{\" u}ven, {\it
Phys.~Rev.~Lett.}~{\bf 51} (1983) 1613.
\bibitem{AE86} P.C.~Aichelburg and
F.~Embacher, {\it Phys.~Rev.}~{\bf D34} (1986) 3006; {\it ibid.}~{\bf
D37}, (1988) 338; {\it ibid.}~{\bf D37}, (1988) 911; {\it ibid.}~{\bf
D37}, (1988) 1436; {\it ibid.}~{\bf D37}, (1988) 2132.}
\bibitem{BKO} R.Brooks, R. Kallosh, and T. Ort\'{\i}n,
 Phys. Rev. D {\bf 52}, 5797-5805 (1995), hep-th/9505116.
\bibitem{Gibb} {G.W.~Gibbons, in {\sl Unified Theories of Elementary
Particles}, Lecture Notes in Physics, Vol. 160, P.~Breitenlohner and
H.P.~D{\" u}rr, eds.,~(Springer, New York, 1982); in {\sl
Supersymmetry,
Supergravity and Related Topics}, proceedings of the XV$^{\rm th}$
GIFT
Seminar, F.~del \'Aguila, {\it et al.},~eds.,~(World Scientific,
Singapore, 1985).}
 {G.W.~Gibbons and C.M.~Hull, {\it Phys.  Lett. }
{\bf
109B} (1982) 190;
C. M. Hull, {\it Commun. Math. Phys.}~{\bf 90} (1983) 545.}
\bibitem{BHGAUGE} R. Kallosh,    Phys. Rev. D {\bf 52}, 6020  (1995),
hep-th/9506113.
\bibitem{KSW} R. Kallosh, M. Shmakova, and W.K. Wong,  Phys. Rev. D {\bf 54},
6284 (1996), hep-th/9607077.
\bibitem{ABCDFFM} L. Andrianopoli, M. Bertolini, A. Ceresole, R. D'Auria, S.
Ferrara, P. Fre,  and T. Magri,   POLFIS-TH-03-96,  hep-th/9605032.
\bibitem{CDF} A. Ceresole, R. D'Auria, and S. Ferrara,   POLFIS-TH-10-95
(1995), hep-th/9509160.
\bibitem{Rey} A similar estimate was made by S.-Y. Rey,   hep-th/9610157.
\bibitem{US}  R.E.~Kallosh, A.~Linde, T.~Ort\'{\i}n,
A.~Peet, and A.~van Proeyen,   Phys.~Rev. D {\bf  46},  5278 (1992),
hep-th/9205027.
\bibitem{KallD} R. Kallosh, ``Bound States of Branes with Minimal Energy,''
Stanford University preprint SU-ITP-96-53 (1996).
\end{references}
\end{document}